\documentclass[sn-nature]{sn-jnl}
\usepackage{anyfontsize}
\usepackage[T1]{fontenc}
\usepackage[utf8]{inputenc}
\usepackage{nicefrac}
\usepackage{graphicx}
\usepackage{hyperref}
\usepackage{support-caption}
\usepackage{subcaption}
\usepackage{multirow}%
\usepackage{amsmath,amssymb,amsfonts}%
\usepackage{amsthm}%
\usepackage{mathrsfs}%
\usepackage[title]{appendix}%
\usepackage{xcolor}%
\usepackage{textcomp}%
\usepackage{manyfoot}%
\usepackage{booktabs}%
\usepackage{bookmark}
\usepackage{algorithm}%
\usepackage{algorithmicx}%
\usepackage{algpseudocode}%
\usepackage{listings}
\usepackage{bm}
\usepackage{tikz}
\usepackage{xcolor}
\usetikzlibrary{arrows,decorations.markings}
\usepackage{csquotes}
\usepackage[skins,theorems,most]{tcolorbox}
\tcbset{highlight math style={enhanced, colframe=black!60!black,colback=white,arc=4pt,boxrule=1pt}} 
\usepackage{soul}
\usepackage{empheq}
\usepackage[normalem]{ulem}
\usepackage{geometry}
\usepackage{natbib}

\definecolor{lime}{HTML}{A6CE39}
\DeclareRobustCommand{\orcidicon}{%
	\begin{tikzpicture}
	\draw[lime, fill=lime] (0,0) 
	circle [radius=0.16] 
	node[white] {{\fontfamily{qag}\selectfont \tiny ID}};
	\draw[white, fill=white] (-0.0625,0.095) 
	circle [radius=0.007];
	\end{tikzpicture}
	\hspace{-2mm}
}
\foreach \x in {A, ..., Z}{%
	\expandafter\xdef\csname orcid\x\endcsname{\noexpand\href{https://orcid.org/\csname orcidauthor\x\endcsname}{\noexpand\orcidicon}}
}
\begin{document}

\title[AArticle Title]{
Some remarks on Frolov-AdS
black hole
surrounded by a fluid of strings }

\newcommand{\orcidauthorA}{0009-0009-7943-5368}
\newcommand{\orcidauthorB}{0000-0001-7893-0265}
\newcommand{\orcidauthorC}{0000-0001-9284-0549}
\newcommand{\orcidauthorD}{0000-0001-5938-1061}
\newcommand{\orcidauthorE}{0000-0003-4515-9245}

\author*[1]{\fnm{F. F.} \sur{Nascimento}\,\orcidA{}}\email{fran.nice.fisica@gmail.com}

\author[1]{\fnm{V. B.} \sur{Bezerra}\,\orcidB{}}\email{valdir@fisica.ufpb.br}

\author[1]{\fnm{J. M.} \sur{Toledo}\,\orcidC{}}\email{jefferson.m.toledo@gmail.com}

\author[2]{\fnm{G. A.} \sur{Marques}\,\orcidD{}}\email{gmarques@df.ufcg.edu.br}

\author[3]{\fnm{J. C.} \sur{Rocha}\,\orcidE{}}\email{julio.rocha@servidor.uepb.edu.br}

\affil*[1]{\orgdiv{Physics Department}, \orgname{Federal University of Para\'iba}, \orgaddress{\city{ Jo\~ao Pessoa}, \postcode{58059-900}, \state{PB}, \country{Brazil}}}

\affil[2]{\orgdiv{Physics Department}, \orgname{Federal University of Campina Grande}, \orgaddress{\city{ Campina Grande}, \postcode{58109-000}, \state{PB}, \country{Brazil}}}

\affil[3]{\orgdiv{Physics Department}, \orgname{State University of Para\'iba}, \orgaddress{\city{Campina Grande}, \postcode{58429-500}, \state{PB}, \country{Brazil}}}


\abstract{A class of new solutions that generalizes the Frolov regular black hole solution is obtained.
The generalization is performed by adding the cosmological constant and surrounding the black hole with a fluid of strings. Among these solutions, some 
preserve the regularity of the original Frolov solution, depending on the values of the parameter $\beta$ 
which labels the different solutions.
A discussion is presented on the 
features of the solutions with respect
to the existence or not of 
singularities, by examining 
the Kretschmann scalar, as well as, by analysing the behavior of the geodesics with 
respect to their completeness. 
It is performed some 
investigations concerning different aspects of thermodynamics,
concerning the role played by the parameter associated to the Frolov regular black
hole solution, as well as, 
of the parameter that codifies 
the presence of the fluid of 
strings.
These are realized by
considering diferent values of the parameter $\beta$, in particular, for $\beta=
- 1/2$, in which case the regularity of the Frolov black hole is 
preserved. All obtained results closely align with the ones obtained by taking the appropriate particularizations.    
 }

\maketitle

\section{Introduction}\label{sec1}

The General Theory of Relativity
predicts the existence of peculiar structures in the Universe which are called black holes. They remained an object of academic interest for a long time, since the discovery of a solution corresponding to a static and spherically symmetric black hole, by Schwarzschild \cite{schwarzschild1916uber} and its generalizations by Reissner and Nordström, with the inclusion of electric charge \cite{reissner1916eigengravitation,nordstrom1918een} and by Kerr and Newman who considered the rotation \cite{kerr1963gravitational,newman1965metric}. It is worth calling attention to the fact that all spacetimes associated with these solutions   have singularities, in which the curvature diverges and the General Theory of Relativity breaks down.
Therefore, the existence of these singular solutions may represent a failure of the theory, which must be corrected, certainly, by constructing a quantum theory of gravity \cite{capozziello2011extended}.

A more direct way to avoid singularities is to construct some models of black holes without spacetime singularities, namely, black holes having regular centers, which are called regular black holes or nonsingular black holes.

A class of black holes without a singularity, namely, regular
or non-singular black holes was constructed by Bardeen \cite{bardeen1968non}, using phenomenological assumptions, which results a metric
 representing a simple redefinition of the mass in the Schwarzschild solution, now depending on the radial coordinate.
Inspired by the solution obtained by Bardeen \cite{bardeen1968non}, in the sequence, different regular
black hole 
solutios
have been obtained 
 in the literature \cite{Dymnikova:1992ux,mars1996models,ayon1998regular,ayon1999new,dymnikova2003spherically,frolov2016notes,sajadi2017nonlinear,hayward2006formation}. The properties of regular black holes were also studied, as, for example, the black hole thermodynamics \cite{saleh2018thermodynamics,molina2021thermodynamics,paul2023more,singh2020thermodynamics}, geodesics \cite{abbas2014geodesic,zhou2012geodesic} and quasinormal modes \cite{fernando2012quasinormal,flachi2013quasinormal,lin2013quasinormal,perez2018region}.

Among the regular spacetimes, in this work, we will focus on the Frolov metric
\cite{frolov2016notes},
which represents a static and spherically symmetric black hole that behaves like a de Sitter spacetime at the origin ($r \rightarrow 0$) and is asymptotically
Reissner-Nordström
for $r \rightarrow \infty$. 
The Frolov solution 
presents an additional charge parameter as compared with 
Hayward
black hole \cite{hayward2006formation}
solution and can be understood as a kind of
generalization
of this one. 
 It can  also be understood as an exact model of a black hole in the General
Theory
of 
Relativity coupled with nonlinear electrodynamics. 

Forty five years ago, Letelier construed a gauge-invariant model of a cloud of strings  \cite{letelier1979clouds}, motivated by the idea that   the fundamental building blocks of of nature are one-dimensional objects, namely,
stings, rather than point particles. Two years later, Letelier generalized the original model, including the pressure. In this context,
instead of a cloud of strings, we have a fluid of strings \cite{letelier1981fluids}. The solution corresponding to a static black hole immersed in this fluid of strings was obtained by Soleng \cite{soleng1995dark}, who argued that this model, in principle, can be used to explain the rotation curves of galaxies.

One of the most important moments in the history of cosmology was the discovery of the fact that the Universe experiences an accelerated expansion \cite{riess1998observational,perlmutter1999measurements,riess1999bvri}. From this accelerated expansion, we can infer that on a large scale, there is a repulsive energy that causes a negative pressure. 
Several mathematical models have been proposed to explain this accelerated phenomenon, among them, the one based on the cosmological constant \cite{copeland2006dynamics}. In the context of astrophysics, the cosmological constant has also been associated with a thermodynamic pressure in black hole systems, which leads to very interesting consequences \cite{caldarelli2000thermodynamics,dolan2011cosmological}.

Since the pioneering studies of Bekenstein and Hawking in the 1970s \cite{bekenstein1973black,hawking1974black,hawking1976black}, the thermodynamics of black holes has been studied, motivated mainly by the belief
that it can connect gravity and quantum mechanics. In these studies, the presence of a cosmological constant has an important role, since, as already mentioned, it can be interpreted as a thermodynamic pressure with a conjugated thermodynamic volume associated with it \cite{caldarelli2000thermodynamics}. Thus, it is possible to study the black hole system using various thermodynamic potentials and calculate the corresponding intensive variables, and additionally to analyze the black hole stability and phase transitions.

In this paper, we obtain a class of solutions that correspond to a generalization of the Frolov black hole solution \cite{frolov2016notes}, in the sense that 
due to the presence of the cosmological constant, which gives rise to an AdS term,  and 
 a fluid of strings.
The resulting spacetime we are calling
Frolov-AdS black hole surrounded by a fluid of strings.  In this scenario, a class of spacetimes is obtained, corresponding to different values of the parameter $\beta$, which codifies the different solutions. 

It is worth calling attention to the fact that the particular solution we have obtained, namely, $\beta=2$, was considered in the literature as a possible solution
that can mimic a perfect fluid dark matter \cite{soleng1995dark,Zhang:2020mxi}.
We perform a detailed study of the black hole thermodynamics, analyzing the system stability and phase transitions.

This paper is organized as follows. In Sec. \ref{sec2}, we obtain a class of solutions corresponding to the Frolov-AdS black hole surrounded by a fluid of string,  behavior, calculate and discuss the results related to the  Kretschmann scalar, and analyse the geodesics concerning their completeness
or incompleteness. In. Sec. \ref{Thermodynamics}, we study some aspects of the thermodynamics of black holes, with emphasis on the behavior of some thermodynamic quantities in which concerns their dependence on the intensity of the fluid of strings. Finally, in Sec. \ref{sec4}, we present our concluding remarks.

\section{Frolov-AdS black hole surrounded by a fluid of strings.}
\label{sec2}
\subsection{Introduction}

Let us start by considering the Frolov
black hole spacetime \cite{frolov2016notes} which can be obtained as a solution of Einstein's equation coupled to a nonlinear electromagnetic field. Thus, taking this solution as a seed, we include the cosmological constant and a fluid of strings surrounding this black hole, and obtain a class of spherically symmetric solutions that generalizes the one obtained by Frolov\cite{frolov2016notes}.

Firstly, let us write the action that describes the system under consideration, taking into account a minimally coupled nonlinear electromagnetic field, given by 
\begin{equation}
S= \frac{1}{16\pi}\int d^4x\sqrt{-g}(R+\mathcal{L}),\label{eq_action}
\end{equation}
with $g$ being the determinant of the metric tensor, $g_{\mu\nu}$, $R$ the scalar curvature and $\mathcal{L}$ the Lagrangian density associated to the nonlinear electromagnetic field \cite{Bronnikov:2000vy}.

Performing the variation of the action given by Eq.(\ref{eq_action}), with respect to the metric, we find the
equations \cite{bronnikov2001regular}

\begin{equation}
G_{\mu\nu} = 2 \left(\frac{\partial\mathcal{L}(F)}{\partial F}F_{\mu\sigma}F^{\sigma}_{\,\,\nu}-\frac{1}{4}g_{\mu\nu}\mathcal{L}(F)\right),
\label{equação de einstein mista}
\end{equation}
where $G_{\mu\nu}=R_{\mu\nu}-\frac{1}{2}Rg_{\mu\nu}$. In which concerns the  nonlinear electromagnetic field, the equations that should be obeyed are the following: 
\begin{gather}
\nabla_\mu\left(\frac{\partial\mathcal{L}(F)}{\partial F}F^{\nu\mu}\right)=0,\\
 \nabla_\mu(*F^{\nu\mu})=0,
\end{gather}
where $F_{\mu\nu} = 2\nabla_{[\mu}A_{\nu]}$ is the nonlinear electromagnetic field, with $A_{\mu}$ being the potential, and  $\mathcal{L}$ is a function of $F=\frac{1}{4}F_{\mu\nu}F^{\mu\nu}$.

The metric of the non-singular (regular) black hole obtained by Frolov \cite{frolov2016notes} is given by

\begin{equation} \label{frolov}
    ds^2 = f(r) dt^2 - f(r)^{-1} dr^2 - r^2 d\Omega^2,
\end{equation}

\noindent where $d \Omega^2 = d\theta^2 + \sin^2 \theta d\phi^2$. The function $f(r)$ is given by

\begin{equation}
   f(r) = 1-\frac{r^2 \left(2 m r-Q^2\right)}{l^2 \left(2 m r+Q^2\right)+r^4},
\label{frolov11}
\end{equation}

\noindent where the additional charge parameter $Q$ characterizes a specific hair and satisfies $0 < Q \leq 1$ and the length below which quantum gravity effects become important satisfies, $l \leq \sqrt{16/27}$ as in the case of the Hayward black hole \cite{hayward2006formation,nascimento2024some2}. In principle, $Q$ admits an interpretation in terms of electric charge measured by an observer situated at infinity, where the metric is asymptotically flat \cite{vagnozzi2023horizon}. In the limit $l\rightarrow 0$, this metric reproduces the Reissner-Nordström metric \cite{reissner1916eigengravitation}.

Using the metric given by Eq. (\ref{frolov11}), we can obtain the following components of the Einstein tensor:

\begin{equation}
G_t^{\;t}=G_r^{\;r}=\frac{l^2 \left(12 m^2 r^2+4 m Q^2 r-3 Q^4\right)+Q^2 r^4}{\left(l^2 \left(2 m r+Q^2\right)+r^4\right)^2},
\label{frolovfseq:1.18}
\end{equation}
\begin{equation}
\begin{aligned}
G_\theta^{\;\theta}=G_\phi^{\;\phi}=&\frac{24 l^2 m^2 r^3 \left(l^2 m-r^3\right)}{\left(l^2 \left(2 m r+Q^2\right)+r^4\right)^3}+\frac{6 l^2 Q^4 r \left(l^2 m+2 r^3\right)}{\left(l^2 \left(2 m r+Q^2\right)+r^4\right)^3}
\\
&-\frac{Q^2 \left(l^4 \left(3 Q^4-28 m^2 r^2\right)+6 l^2 m r^5+r^8\right)}{\left(l^2 \left(2 m r+Q^2\right)+r^4\right)^3},
\end{aligned}
\label{frolovfseq:1.19}
\end{equation}

\noindent which, according to Einstein's equations, are proportional to the energy-momentum tensor of the source.

Now, let us generalize the metric obtained by Frolov \cite{frolov2016notes}, given by
Eqs.(\ref{frolov11}), taking into account the cosmological constant and the fluid of strings. To do this, we add the term $-\Lambda g_{\mu\nu}$ and the energy momentum tensor corresponding to the fluid of strings 
to the lhs and rhs, respectively of Eq. (\ref{equação de einstein mista}). Thus, we can write

\begin{equation}
R_{\mu\nu}-\frac{1}{2}Rg_{\mu\nu}-\Lambda g_{\mu\nu}= 2 \left(\frac{\partial\mathcal{L}(F)}{\partial F}F_{\mu\sigma}F^{\sigma}_{\,\,\nu}-\frac{1}{4}g_{\mu\nu}\mathcal{L}(F)\right)+ T_{\mu\nu}^{\text{FS}},\label{geral_eq}
\end{equation}
where $T_{\mu\nu}^{\text{FS}}$ refers to the energy-momentum tensor of the fluid of strings. Therefore, the contents of the rhs represent an effective energy-momentum tensor, with the first arising from the nonlinear electromagnetic field, while the second refers to the fluid of strings.

{
\subsection{Energy-momentum
tensor of a fluid of strings}

The trajectory of a  particle
which
moves with a
four-velocity  $u^{\mu}=dx^\mu/d\lambda$, where $\lambda$ is a parameter, can be described by the curve
$x=x(\lambda)$. In the case of a moving infinitesimally thin string, the trajectory corresponds to a two-dimensional world sheet  described by \cite{Letelier:1979ej}
\begin{equation}
x^{\mu}=x^{\mu}(\lambda^{a}), \; \;a=0,1,
\label{eq:1.10}
\end{equation}
with $\lambda_0$ and $\lambda_1$ being timelike and spacelike parameters, respectively. 
We can associate with this world sheet, a bivector
$\Sigma^{\mu\nu}$, 
given by
\cite{Letelier:1979ej}
\begin{equation}
\Sigma^{\mu\nu}=\epsilon^{ab}\frac{\partial{x^\mu}}{\partial{\lambda^{a}}}\frac{\partial{x^\nu}}{\partial{\lambda^{b}}},
\label{eq:1.11}
\end{equation}
where $\epsilon^{ab}$ is the two-dimensional Levi-Civita symbol, in
which 
the following values will be considered: $\epsilon^{01}=-\epsilon^{10}=1$.

It is worth emphasizing that on this world sheet, there will be an induced metric, $\gamma_{ab}$, with $a, b = 0, 1$, such that,
\begin{equation}
\gamma_{ab}=g_{\mu\nu}\frac{\partial{x^\mu}}{\partial{\lambda^{a}}}\frac{\partial{x^\nu}}{\partial{\lambda^{b}}},
\label{eq:1.12}
\end{equation}
whose determinant
is denoted by $\gamma$. 

The energy-momentum tensor associated with a dust cloud is given by $T^{\mu\nu}=\rho u^{\mu}u^{\nu}$, with $u^{\mu}$ being the normalized four-velocity
of a dust particle and $\rho$ the proper density of the flow. 
Analogously, for a cloud of strings, the energy-momentum tensor can be defined as \cite{Letelier:1979ej}
\begin{equation}
T^{\mu\nu}=\rho\frac{\Sigma^{\mu\beta}\Sigma_{\beta}^{\;\nu}}{(-\gamma)^{1/2}},
\label{eq:1.13}
\end{equation}
where $\gamma=\frac{1}{2}\Sigma^{\mu\nu}\Sigma_{\mu\nu}$.

Let us consider a more
general case, taking into account the pressure of a fluid of strings. 
This scenario corresponds to a  
perfect fluid of strings with pressure $p$, whose energy-momentum tensor is given by \cite{letelier1981}
\begin{equation}
T^{\mu\nu}=(p+\sqrt{-\gamma}\rho)\frac{\Sigma^{\mu\beta}\Sigma_\beta^{\;\nu}}{(-\gamma)}+p q^{\mu\nu}.
\label{eq:1.14}
\end{equation}

Taking into account the energy-momentum tensor given by Eq. (\ref{eq:1.14}), it was obtained the metric corresponding to a static black hole surrounded by a fluid of strings \cite{Soleng:1993yr}. In this context, it was assumed that the components of the energy-momentum tensor are related through the equations
\begin{equation}
T_{t}^{\;t}= T_{r}^{\;r},
\label{eq:1.15}
\end{equation}
\begin{equation}
T_{t}^{\;t}= -\beta T_{\theta}^{\;\theta}= -\beta T_{\phi}^{\;\phi},
\label{eq:1.16}
\end{equation}
where $\beta$ is a dimensionless constant. The energy-momentum tensor, whose components are given by Eqs. (\ref{eq:1.15})-(\ref{eq:1.16}), was interpreted as being associated with a kind of anisotropic fluid with spherical symmetry \cite{Dymnikova:1992ux, soleng1994correction}. 
\par
It is worth calling attention to the fact that this kind of energy-momentum tensor has been used in different scenarios \cite{Salgado:2003ub,Giambo:2002wr,Dymnikova:2001fb}. In particular, in Ref. \cite{Salgado:2003ub} it is shown that the conditions imposed by Eq. (\ref{eq:1.16}) permit to obtaining of a class of spherically symmetric solutions of Einstein’s field equations with two parameters.
\par
In this paper, we will consider the components of the energy-momentum tensor for the fluid of strings obtained in \cite{Toledo:2020xxc} and already used in a previous paper\cite{nascimento2024some}.  These components are given by \cite{Toledo:2020xxc}

\begin{equation}
T_{t}^{\;t}= T_{r}^{\;r}= -\frac{\epsilon}{r^2}\left(\frac{b}{r}\right)^{2/\beta},
\label{eq:1.17}
\end{equation}
\begin{equation}
T_{\theta}^{\;\theta}= T_{\phi}^{\;\phi}= \frac{\epsilon}{\beta r^2}\left(\frac{b}{r}\right)^{2/\beta},
\label{eq:1.18}
\end{equation}
where $b$ is a positive integration constant and $\epsilon=\pm1$, with the $\pm$ referring to the signs of the energy density of the fluid of strings.
}
%
%

\subsection{Frolov black hole solution with cosmological constant and surrounded by a fluid of strings}

Now, let us consider Eq. (\ref{geral_eq}) with the 
components of the energy-momentum tensor of the fluid of strings
 given by Eqs. (\ref{eq:1.17}) and  (\ref{eq:1.18}). 

The line element for a static and spherically symmetric spacetime can be written as:
\begin{equation}
ds^2=e^{\nu(r)} dt^2-e^{\lambda(r)} dr^2-r^2 d\theta^2-r^2\sin^2\theta d\phi^2.
\label{eq:line_element}
\end{equation}

Einstein's field equations for the present
case, in which the presence of the  cosmological constant and of the fluid of strings
are taken into account, can be written as 

\begin{equation}
e^{-\lambda}\left(\frac{\lambda'}{r}-\frac{1}{r^2}\right)+\frac{1}{r^2}-\Lambda=\frac{l^2 \left(12 m^2 r^2+4 m Q^2 r-3 Q^4\right)+Q^2 r^4}{\left(l^2 \left(2 m r+Q^2\right)+r^4\right)^2}-\frac{\epsilon}{r^2}\left(\frac{b}{r}\right)^{2/\beta},
\label{frolovfseq17}
\end{equation}

\begin{equation}
-e^{-\lambda}\left(\frac{\nu'}{r}+\frac{1}{r^2}\right)+\frac{1}{r^2}-\Lambda=\frac{l^2 \left(12 m^2 r^2+4 m Q^2 r-3 Q^4\right)+Q^2 r^4}{\left(l^2 \left(2 m r+Q^2\right)+r^4\right)^2}-\frac{\epsilon}{r^2}\left(\frac{b}{r}\right)^{2/\beta},
\label{frolovfseq18}
\end{equation}

\begin{equation}
\begin{aligned}
\frac{1}{2}e^{-\lambda}\left(\frac{\nu'\lambda'}{2}+\frac{\lambda'}{r}-\frac{\nu'}{r}-\frac{\nu'^2}{2}-\nu''\right)-\Lambda=&+\frac{24 l^2 m^2 r^3 \left(l^2 m-r^3\right)}{\left(l^2 \left(2 m r+Q^2\right)+r^4\right)^3}
\\
&+\frac{6 l^2 Q^4 r \left(l^2 m+2 r^3\right)}{\left(l^2 \left(2 m r+Q^2\right)+r^4\right)^3}
\\
&-\frac{Q^2 \left(l^4 \left(3 Q^4-28 m^2 r^2\right)+6 l^2 m r^5+r^8\right)}{\left(l^2 \left(2 m r+Q^2\right)+r^4\right)^3}
\\
&+\frac{\epsilon}{\beta r^2}\left(\frac{b}{r}\right)^{2/\beta}.
\end{aligned}
\label{frolovfseq19}
\end{equation}

\noindent From Eqs. (\ref{frolovfseq17}) and (\ref{frolovfseq18}), it follows immediately that:

\begin{equation}
\nu=-\lambda + const.
\label{frolovfseq20}
\end{equation}
The constant can be absorbed into a rescaling of the time coordinate, so that,

\begin{equation}
\nu=-\lambda.
\label{frolovfseq20}
\end{equation}
\noindent Adding Eqs. (\ref{frolovfseq17}) and (\ref{frolovfseq18}) and considering Eq. (\ref{frolovfseq20}), after some algebraic manipulations, we obtain:

\begin{equation}
e^{-\lambda}\frac{\lambda'}{r}-e^{-\lambda}\frac{1}{r^2}+\frac{1}{r^2}-\Lambda=\frac{l^2 \left(12 m^2 r^2+4 m Q^2 r-3 Q^4\right)+Q^2 r^4}{\left(l^2 \left(2 m r+Q^2\right)+r^4\right)^2}-\frac{\epsilon}{r^2}\left(\frac{b}{r}\right)^{2/\beta}.
\label{frolovfseq21}
\end{equation}

\noindent Then, consider an arbitrary function $f(r)$,
such that

\begin{equation}
\nu=-\lambda=ln(1+f(r)).
\label{frolovfseq22}
\end{equation}

\noindent Taking  Eqs. (\ref{frolovfseq20}) and (\ref{frolovfseq22}) into account, we can write Eqs. 
(\ref{frolovfseq21}) and
(\ref{frolovfseq19}) and, respectively, as follows:

\begin{equation}
-\frac{1}{r^2}(rf'+f)-\Lambda=\frac{l^2 \left(12 m^2 r^2+4 m Q^2 r-3 Q^4\right)+Q^2 r^4}{\left(l^2 \left(2 m r+Q^2\right)+r^4\right)^2}-\frac{\epsilon}{r^2}\left(\frac{b}{r}\right)^{2/\beta},
\label{frolovfseq25}
\end{equation}

\begin{equation}
\begin{aligned}   
2\frac{f'}{r}+f''+2\Lambda=&-\frac{48 l^2 m^2 r^3 \left(l^2 m-r^3\right)}{\left(l^2 \left(2 m r+Q^2\right)+r^4\right)^3}
\\
&-\frac{12 l^2 Q^4 r \left(l^2 m+2 r^3\right)}{\left(l^2 \left(2 m r+Q^2\right)+r^4\right)^3}
\\
&+\frac{2Q^2 \left(l^4 \left(3 Q^4-28 m^2 r^2\right)+6 l^2 m r^5+r^8\right)}{\left(l^2 \left(2 m r+Q^2\right)+r^4\right)^3}
\\
&-\frac{2\epsilon}{\beta r^2}\left(\frac{b}{r}\right)^{2/\beta}.
\end{aligned}
\label{frolovfseq26}
\end{equation}

\noindent 
Multiplying both
Eqs. (\ref{frolovfseq25}) and (\ref{frolovfseq26}) by $r^2$ and adding the obtained equations, we get the following differential equation:

\begin{equation}
\begin{aligned}  
&r^2f''+rf'-f+\Lambda r^2+\epsilon\left(\frac{\beta+2}{\beta}\right)\left(\frac{b}{r}\right)^{2/\beta}
\\
&-\frac{l^2 r^2\left(12 m^2 r^2+4 m Q^2 r-3 Q^4\right)+Q^2 r^6}{\left(l^2 \left(2 m r+Q^2\right)+r^4\right)^2}
\\
& +\frac{48 l^2 m^2 r^5 \left(l^2 m-r^3\right)}{\left(l^2 \left(2 m r+Q^2\right)+r^4\right)^3}+\frac{12 l^2 Q^4 r^3 \left(l^2 m+2 r^3\right)}{\left(l^2 \left(2 m r+Q^2\right)+r^4\right)^3}
\\
&-\frac{2Q^2 r^2\left(l^4 \left(3 Q^4-28 m^2 r^2\right)+6 l^2 m r^5+r^8\right)}{\left(l^2 \left(2 m r+Q^2\right)+r^4\right)^3}=0,
\end{aligned}
\label{frolovfseq27}
\end{equation}

\noindent whose solution is given by:

\begin{equation}
f(r)=-\frac{r^2 \left(2 m r-Q^2\right)}{l^2 \left(2 m r+Q^2\right)+r^4}-\frac{\Lambda  r^2}{3}+
\left\{\begin{array}{rcl}\epsilon b[1+2\log (r)]/2r \text{ for } \beta=2,\\
\epsilon\beta(\beta -2)^{-1} \left(\frac{b}{r}\right)^{2/\beta} \text{ for }\beta\neq 2.
\end{array}
\right.
\label{frolovfseq28}
\end{equation}

\noindent So, substitute Eq. (\ref{frolovfseq28}) into Eq. (\ref{frolovfseq22}) and then into Eq. (\ref{eq:line_element}), we finally obtain the Frolov-AdS black hole surrounded by a fluid of strings, where the line element can be written as

\begin{equation}
ds^2=f(r)dt^2-f(r)^{-1}dr^2-r^2 d\Omega^2,
\label{frolovfseq29}
\end{equation}

\noindent where $d\Omega^2=d\theta^2+\sin^2\theta d\phi^2$. The function $f(r)$ is given by:

\begin{equation}
f(r)=1-\frac{r^2 \left(2 m r-Q^2\right)}{l^2 \left(2 m r+Q^2\right)+r^4}-\frac{\Lambda  r^2}{3}+
\left\{\begin{array}{rcl}\epsilon b[1+2\log (r)]/2r\text{ for }\beta=2,\\
\epsilon\beta(\beta -2)^{-1} \left(\frac{b}{r}\right)^{2/\beta}\text{ for }\beta\neq 2.
\end{array}
\right.
\label{frolovfseq30}
\end{equation}

Note that we can recover some other solutions from this metric if we make the choices displayed in the Table (\ref{table_dif_ST}).
\begin{table}[ht]
\caption{Special cases for $l=Q=\Lambda=0$.}
\begin{tabular}{lcclccclcccclcl}
\hline
&$\epsilon$ & & & & $b$ & & & & $\beta$ & & & & Spacetime &\\\hline
&$1$ & & & & $\mathbb{R}^{*}$ & & & & $lim_{\beta\rightarrow\infty}$& & & & Letelier &\\
&$1$ & & & & $Q$ & & & & $1$ & & & & Reissner-Nordström &\\
&$1$ & & & & $\Lambda^{-1/2}$& & & & $-1$ & & & & de Sitter&\\\hline
\end{tabular}
\label{table_dif_ST}
\end{table}

Note that the class of solutions obtained can be considered as sourced by two fluids, namely, one corresponding to the fluid of strings
and the other to a
nonlinear electromagnetic
field. In addition, a cosmological constant is considered.

\subsection{The
Kretschmann scalar: calculation and discussion}

Now, let us discuss the existence of physical singularities of spacetime
by considering one of the scalars
constructed with the curvature tensor, namely the Kretschmann scalar. It
is worth
pointing out that the absence of singularities in such a curvature scalar
does not guarantee that the spacetime is regular.
The discussions will be compared with
similar ones for a Frolov black hole, which is regular.
In the section that follows, we will analyze the same question, but this time, taking into account the geodesics \cite{hawking2023large,wald2010general}.

In what follows, we will determine and analyze the limits of the Kretschmann scalar when $r\rightarrow 0$ and $r\rightarrow \infty$, for $l>0$, $Q>0$, $b>0$ and $\Lambda$ and $\epsilon\in \mathbb{R}$, for some values of $\beta$.
\begin{itemize}

\item {For $\beta=-2$}, the Kretschmann scalar diverges very close to the origin and is finite in a region very far from the black hole. 

\begin{equation}
\lim_{r\rightarrow 0}K=\infty.
\end{equation}
 
\begin{equation}
\lim_{r\rightarrow \infty}K=\frac{8 \Lambda ^2}{3}.
\end{equation}
\item 
For $\beta=-1$, the Kretschmann scalar is finite, and is given by

\begin{equation}
\lim_{r\rightarrow 0}K=\frac{8 \left(b^2 \left(3-\Lambda  l^2\right)+l^2 \epsilon \right)^2}{3 b^4 l^4}.
\end{equation}
\begin{equation}
\lim_{r\rightarrow \infty}K=\frac{8 \left(\epsilon -b^2 \Lambda \right)^2}{3 b^4}.
\end{equation}
 
\item {For $-1<\beta <0$}, the Kretschmann scalar is finite, close to the origin, and diverges for points very far from the black hole, according to 

\begin{equation}
\lim_{r\rightarrow 0}K=\frac{8 \left(\Lambda  l^2-3\right)^2}{3 l^4}.
\end{equation}
\begin{equation}
\lim_{r\rightarrow \infty}K=\infty.
\end{equation}
 
\item {For $\beta=1$, $\beta=2$ and $\beta=3$}, the Kretschmann scalar diverges very close to the origin and has a finite value for points very far from the black hole.

\begin{equation}
\lim_{r\rightarrow 0}K=\infty.
\end{equation}
\begin{equation}
\lim_{r\rightarrow \infty}K=\frac{8 \Lambda ^2}{3}.
\end{equation}
\end{itemize}

These results tell us that the inclusion of the fluid of strings changes the regularity of the Frolov solution for the following values of $\beta$, such that $\beta < -1$ and $\beta > 0$. Otherwise, the regularity of the Frolov black hole solution is preserved in the interval $-1\leq\beta<0$.

\subsection{Geodesics and Effective Potential}

Let us now consider the static and spherically symmetric solution given by Eq. (\ref{frolovfseq30})
 and analyze the geodesic equations. To do this, consider the Lagrangian $\mathcal{L}$, such that

\begin{equation}
    2\mathcal{L} = f(r) \dot{t}^2-\frac{1}{f(r)}\dot{r}^2-r^2\dot{\theta}^2-r^2\sin^2\theta\dot{\phi}^2.\label{lagrangianfrolovfs}
\end{equation}
The “point” represents the derivative concerning proper time, $\tau$. For simplicity, let's restrict the analysis of the geodesics to the equatorial plane of the black hole, $\theta=\frac{\pi}{2}$. Using the Euler-Lagrange equations, we obtain
\begin{align}
    E &= f(r)\dot{t},
\label{energiafrolovfs}
\end{align}
\begin{align}
    J &= -r^2\dot{\phi},
\label{momento angularfrolovfs}
\end{align}

\noindent where $E$ and $J$ are constants of motion.
which can be interpreted as the energy $E$ and the angular momentum $J$ of the particle moving around the black hole. 

By rescaling the parameter $\tau$, we can define $L=2\mathcal{L}$, which, for time-like geodesics, is equal to $+1$, for space-like geodesics is equal to $-1$, and is equal to $0$ for null geodesics \cite{chandrasekhar1983mathematical}. Substituting Eqs. (\ref{energiafrolovfs}) and (\ref{momento angularfrolovfs}) into Eq. (\ref{lagrangianfrolovfs}), we get

\begin{equation}
\dot{r}^2=E^2-V_{eff},
\label{energia e potencial efetivofrolovfs}
\end{equation}

\noindent where

\begin{equation}
V_{eff}=f(r)\left(\frac{J^2}{r^2}+L\right).
\label{potenciale efetivofrolovfs}
\end{equation}

Let's now consider the problem of the massive particle $(L=1)$ falling radially $(J=0)$ into the black hole. The equation of radial geodesic motion of this test particle is given by

\begin{equation}
    \dot{r}^2 = E^2- f(r),
\end{equation}
while the effective potential is as follows
\begin{equation}
    V_{eff} = f(r).
\end{equation}

\begin{figure}[h!]
  \centering
  \begin{subfigure}[b]{0.45\textwidth}
    \includegraphics[width=\textwidth]{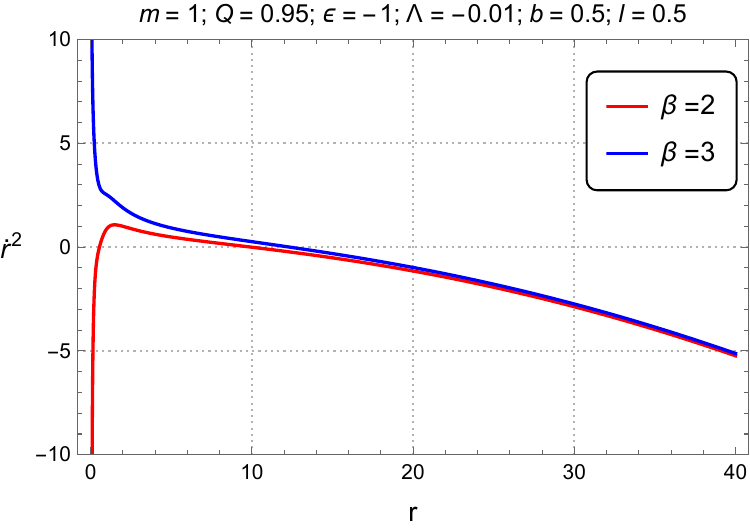}
    \caption{}\label{afrolovfs}
  \end{subfigure}
  \vspace{.5cm}
  \begin{subfigure}[b]{0.47\textwidth}
    \includegraphics[width=\textwidth]{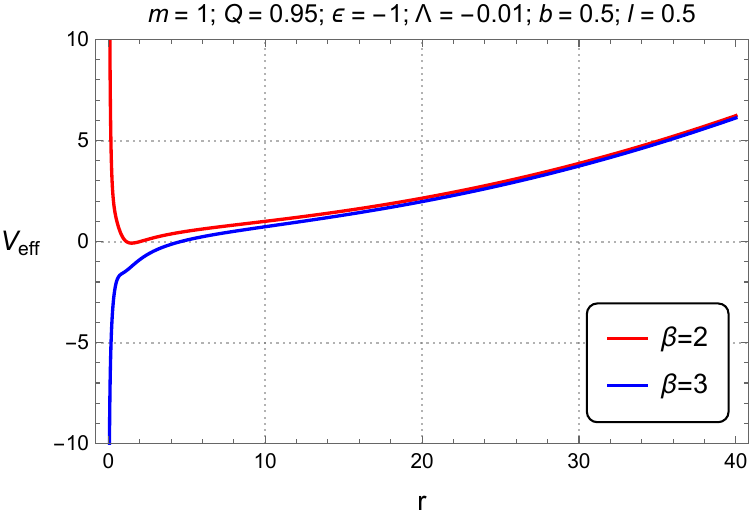}
    \caption{}\label{bfrolovfs}
  \end{subfigure}
   \begin{subfigure}[b]{0.45\textwidth}
    \includegraphics[width=\textwidth]{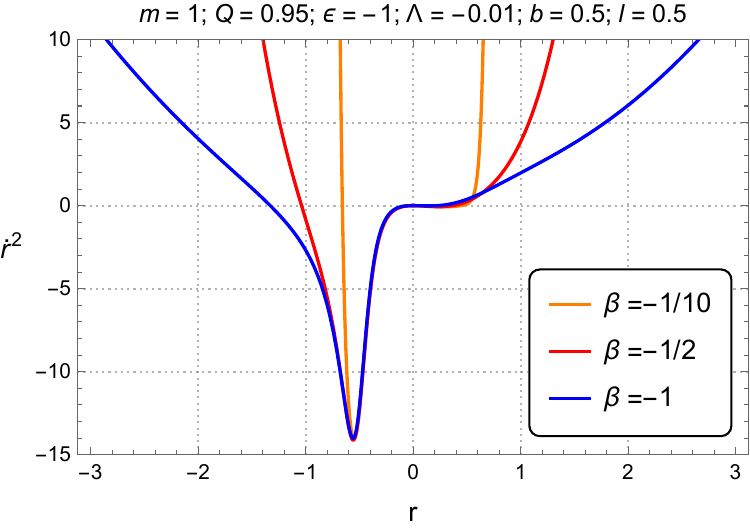}
    \caption{}\label{cfrolovfs}
  \end{subfigure}
   \begin{subfigure}[b]{0.47\textwidth}
    \includegraphics[width=\textwidth]{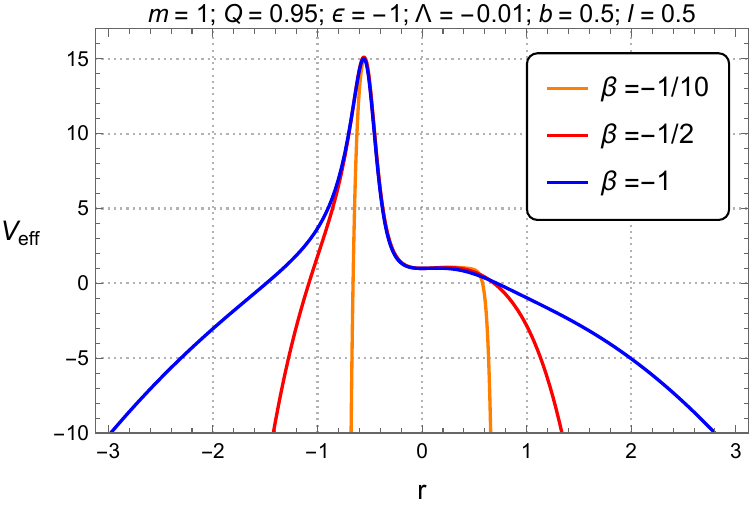}
    \caption{}\label{dfrolovfs}
  \end{subfigure}
 \caption{The graph on the left shows $\dot{r}^2$ for $E=2$. The graph on the right shows the effective potential $V_{eff} = f(r)$.}
  \label{testandofrolovfs}
\end{figure}

The graphs for $\dot{r}^2$ and $V_{eff}$ are shown in Fig. \ref{testandofrolovfs}.

Let's now compare the results predicted by the values of the Kretschmann scalar when $r\rightarrow 0$, with the characteristics of the geodesics, namely whether they are complete or incomplete.

In Fig. \ref{bfrolovfs}, we can conclude that the particle cannot reach the point $r=0$ in a finite time, for both values of $\beta$, namely, $\beta=2$ and $\beta=3$. This indicates that the geodesics are incomplete and therefore space-time is singular. This conclusion is confirmed by the fact that, for $\beta=2$, the Kretschmann scalar is infinite when $r\rightarrow 0$.

 For
 $\beta=-1/10$,
$\beta=-1/2$ and $\beta=-1$ in Fig. \ref{dfrolovfs}, the test particle manages to cross the potential barrier and reach the point $r=0$ in a finite time. This indicates that the geodesics are complete and therefore space-time is regular.

In summary, the analysis of the geodesics to know the characteristics of the spacetimes in relation to the singularity confirms what is predicted by the results provided by the Kretschmann scalar, at least in the cases considered.

\section{Black Hole Thermodynamics}
\label{Thermodynamics}

In this section, we will study the thermodynamics of the Frolov-AdS black hole with a fluid of strings, examining the behavior of mass, Hawking temperature, and heat capacity as a function of entropy.

It is worth calling attention to two considerations that we will assume in this work: 

(i) We will study the solution $\beta=2$ since this choice may represent the presence of dark matter \cite{soleng1995dark};
\par
(ii) We will also analyze thermodynamics for values of $-1\leq\beta<0$, since in this interval we obtain regular solutions, preserving the main characteristic of the Frolov solution, namely regularity.
\subsection{Black hole mass}

Let $r_h$ be the radius of the horizon, so we have $f(r_h)=0$, where $f(r)$ is given by Eq. (\ref{frolovfseq30}). So we can write the mass of the black hole in terms of $r_h$ using the following equation:

\begin{equation}
\begin{aligned}  
m^{(\beta=2)}=\frac{l^2 Q^2 \left(3 b \epsilon -2 \Lambda  r_h^3+6 r_h\right)+6 b \epsilon  \log \left(r_h\right) \left(r_h^4+l^2 Q^2\right)+r_h^3 \left(3 b \epsilon  r_h-2 \Lambda  r_h^4+6 r_h^2+6 Q^2\right)}{2 r_h \left(l^2 \left(-3 b \epsilon +2 \Lambda  r_h^3-6 r_h\right)-6 b l^2 \epsilon  \log \left(r_h\right)+6 r_h^3\right)},
\label{eq:1.59frolovfluido}
\end{aligned}
\end{equation}

\begin{equation}
\begin{aligned}  
m^{(\beta\neq2)}=\frac{3 \beta  \epsilon  \left(\frac{b}{r_h}\right){}^{2/\beta } \left(r_h^4+l^2 Q^2\right)+(\beta -2) \left(l^2 Q^2 \left(3-\Lambda  r_h^2\right)+3 r_h^2 \left(r_h^2+Q^2\right)-\Lambda  r_h^6\right)}{2 l^2 r_h \left((\beta -2) \left(\Lambda  r_h^2-3\right)-3 \beta  \epsilon  \left(\frac{b}{r_h}\right){}^{2/\beta }\right)+6 (\beta -2) r_h^3},
\label{eq:1.59frolovfluido11}
\end{aligned}
\end{equation}

\noindent written in terms of the parameter that describes the presence of the fluid of strings, $b$. Note that if $b=0$, we recover the mass of the regular Frolov-AdS black hole, without the fluid of strings, in terms of the radius of the horizon. Considering $l=0$, $\Lambda=0$ and $b=0$, we recover the mass of the Reissner-Nordström black hole.

The expression for the mass parameter can be rewritten in terms of the entropy,
by using the following relation \cite{bekenstein1973black} 

\begin{equation}
S=\frac{A}{4}=\pi r_h^2,
\label{eq:1.61frolovfluido}
\end{equation}as follows

\begin{equation}
\begin{aligned} 
m^{(\beta=2)}=&[\pi ^2 l^2 Q^2 \left(-3 \pi ^{3/2} b \epsilon +2 \Lambda  S^{3/2}-6 \pi  \sqrt{S}\right)
\\
&+S^{3/2} \left(-3 \pi ^{3/2} b \sqrt{S} \epsilon -6 \pi ^2 Q^2+2 \Lambda  S^2-6 \pi  S\right)
\\
&
-3 \pi ^{3/2} b \epsilon  \log \left(\frac{S}{\pi }\right) \left(\pi ^2 l^2 Q^2+S^2\right)]/
\\
&[2 \pi ^{3/2} \sqrt{S} \left(l^2 \left(3 \pi ^{3/2} b \epsilon -2 \Lambda  S^{3/2}+6 \pi  \sqrt{S}\right)+3 \pi ^{3/2} b l^2 \epsilon  \log \left(\frac{S}{\pi }\right)-6 S^{3/2}\right)],
\label{eq:1.62frolovfluido}
\end{aligned}
\end{equation}

\begin{equation}
\begin{aligned} 
m^{(\beta\neq2)}=\frac{3 \pi ^{1/\beta } \beta  \epsilon  \left(\frac{b}{\sqrt{S}}\right)^{2/\beta } \left(l^2 Q^2+\frac{S^2}{\pi ^2}\right)+(\beta -2) \left(l^2 Q^2 \left(3-\frac{\Lambda  S}{\pi }\right)+\frac{3 S \left(Q^2+\frac{S}{\pi }\right)}{\pi }-\frac{\Lambda  S^3}{\pi ^3}\right)}{\frac{2 l^2 \sqrt{S} \left((\beta -2) \left(\frac{\Lambda  S}{\pi }-3\right)-3 \pi ^{1/\beta } \beta  \epsilon  \left(\frac{b}{\sqrt{S}}\right)^{2/\beta }\right)}{\sqrt{\pi }}+\frac{6 (\beta -2) S^{3/2}}{\pi ^{3/2}}}.
\label{eq:1.62frolovfluido11}
\end{aligned}
\end{equation}

\begin{figure}[h!]
  \centering
  \begin{subfigure}[b]{0.45\textwidth}
    \includegraphics[width=\textwidth]{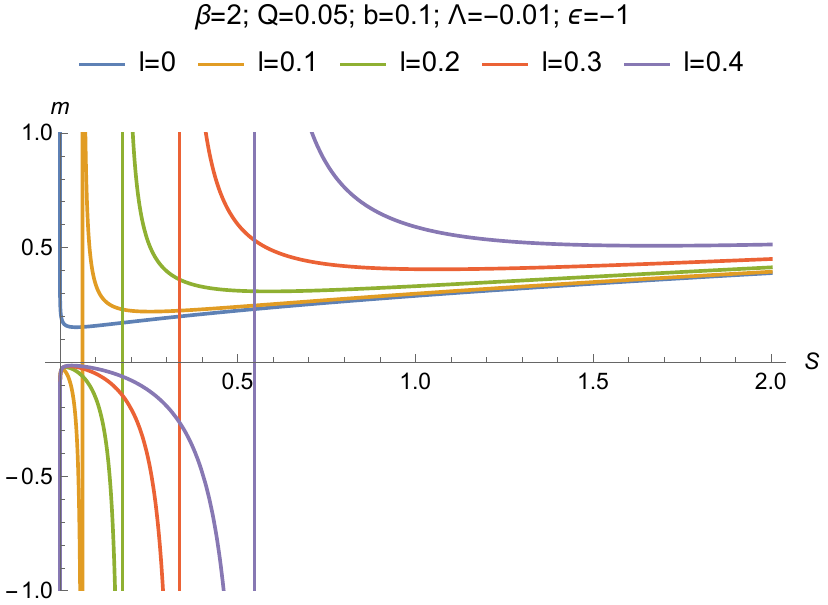}
    \caption{}\label{im2afrolovfluido}
  \end{subfigure}
  \vspace{.5cm}
  \begin{subfigure}[b]{0.47\textwidth}
    \includegraphics[width=\textwidth]{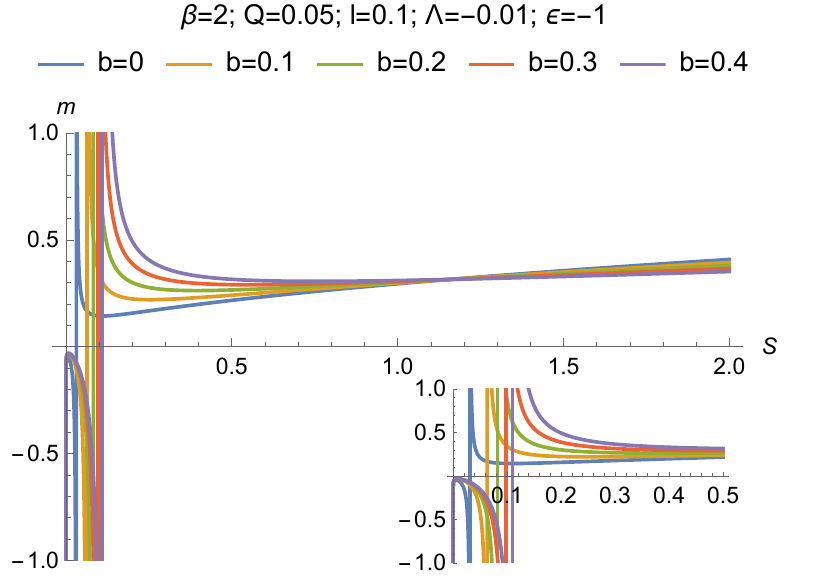}
    \caption{}\label{im2bfrolovfluido}
  \end{subfigure}
   \begin{subfigure}[b]{0.45\textwidth}
    \includegraphics[width=\textwidth]{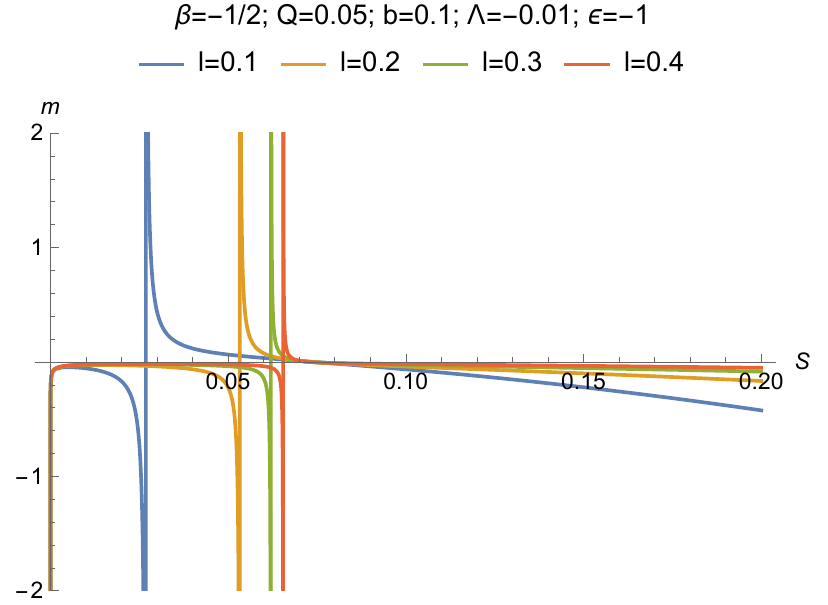}
    \caption{}\label{ffc1}
  \end{subfigure}
   \begin{subfigure}[b]{0.47\textwidth}
    \includegraphics[width=\textwidth]{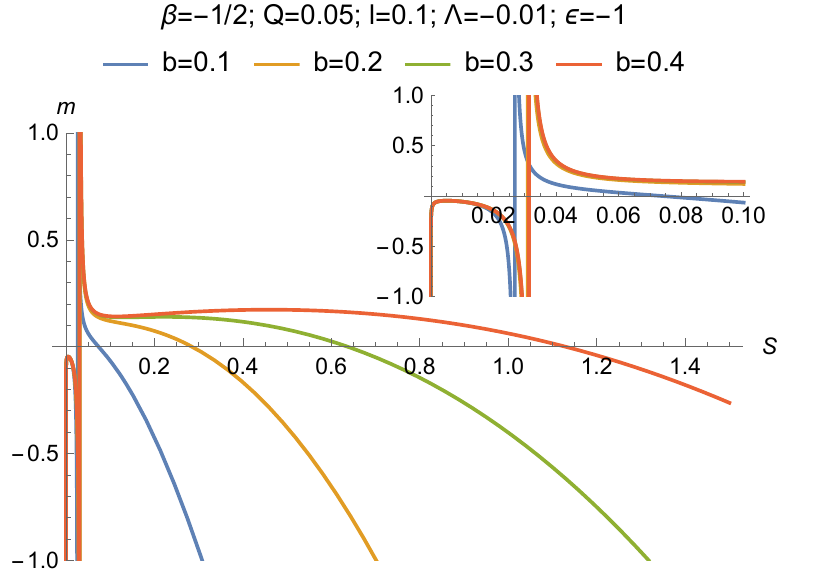}
    \caption{}\label{ffc2}
  \end{subfigure}
 \caption{Black hole mass as a function of entropy $m(S)$ for different values of $b$ and $l$.}
  \label{imfrolovfluido1}
\end{figure}

In Figs. \ref{im2afrolovfluido}-\ref{ffc2}, we represent the behavior of the mass parameter, $m$, as a function of the entropy of the black hole, $S$, in different situations. Note that Figs. \ref{im2afrolovfluido}-\ref{im2bfrolovfluido}, describe the behavior of mass as a function of entropy for the singular Frolov-AdS spacetime with dark matter fluid ($\beta=2$), while Figs. \ref{ffc1}-\ref{ffc2}, describe the mass as a function of entropy for the regular Frolov-AdS spacetime with fluid of strings for $\beta=-1/2$. Note also that, for the Frolov-AdS black hole with fluid of strings ($\beta=2$ and $\beta=-1/2$), the mass parameter has positive and negative values depending on the parameters of the black hole. 

It is worth emphasizing that Figs. \ref{im2afrolovfluido}-\ref{im2bfrolovfluido} have the same limits, so that the mass will show positive values as the entropy takes on increasing values. If we set $S=2$ in Figs. \ref{im2afrolovfluido}-\ref{im2bfrolovfluido}, we can see that the mass increases as $l$ increases and $b$ decreases.
In Figs. \ref{ffc1}-\ref{ffc2}, the limits are the same, but opposite to those found in Figs. \ref{im2afrolovfluido}-\ref{im2bfrolovfluido}. If we set $S=0.2$ in Figs. \ref{ffc1}-\ref{ffc2}, we conclude that the mass increases as we increase the values of $l$ and $b$.

\subsection{Hawking temperature}

Firstly, let us calculate the surface gravity, $\kappa$, for the space-timesunder consideration, defined as: 

\begin{equation}
\kappa=\frac{f'(r)}{2}\left|\frac{}{}_{r_{h}}\right.,
\label{eq:1.63frolovfluido}
\end{equation}

\noindent where $(')$ means the derivative concerning the radial coordinate. Using surface gravity, the Hawking temperature for a stationary space-time is given by \cite{hawking1975particle}:

\begin{equation}
T_\kappa=\frac{\kappa}{2\pi}.
\label{eq:1.64frolovfluido}
\end{equation}

Now, consider
$f(r)$ given by Eq.(\ref{frolovfseq30}) and substitute into
Eqs.
(\ref{eq:1.63frolovfluido})
and
(\ref{eq:1.64frolovfluido}). 
Thus, the Hawking temperature, $T_\kappa=T$, for the Frolov-AdS black hole with a fluid of strings can be written as:

\begin{equation}
\begin{aligned}
 T^{(\beta=2)}=&\frac{1}{144 \pi ^{7/2} S^{5/2} \left(2 l^2 Q^2+\frac{S^2}{\pi ^2}\right)}\left[36 \pi ^4 l^4 Q^2 S-12 \Lambda  S^5 \left(\Lambda  l^2+3\right)+36 \pi  S^4 \left(2 \Lambda  l^2+1\right)\right.
 \\
 &\left.+4 \pi ^2 S^3 \left(\Lambda ^2 l^4 Q^2-27 l^2-9 Q^2\right)+36 \pi ^{3/2} b S^{7/2} \epsilon  \left(\Lambda  l^2+2 \Lambda  l^2 \log \left(\frac{\sqrt{S}}{\sqrt{\pi }}\right)+1\right)\right.
\\
&\left.-12 \pi ^{7/2} b l^2 Q^2 S^{3/2} \epsilon  \left(\Lambda  l^2+2 \left(\Lambda  l^2+9\right) \log \left(\frac{\sqrt{S}}{\sqrt{\pi }}\right)+3\right)\right.
\\
&\left.-3 \pi ^3 l^2 S^2 \left(36 b^2 \epsilon ^2 \log ^2\left(\frac{\sqrt{S}}{\sqrt{\pi }}\right)+36 b^2 \epsilon ^2 \log \left(\frac{\sqrt{S}}{\sqrt{\pi }}\right)+9 b^2 \epsilon ^2+8 Q^2 \left(\Lambda  l^2+6\right)\right)\right.
\\
&\left.+36 \pi ^{9/2} b l^4 Q^2 \sqrt{S} \epsilon  \left(\log \left(\frac{S}{\pi }\right)+1\right)\right.
\\
&\left.+9 \pi ^5 b^2 l^4 Q^2 \epsilon ^2 \left(\log \left(\frac{S}{\pi }\right)+1\right)^2-108 \pi ^{5/2} b l^2 S^{5/2} \epsilon  \left(\log \left(\frac{S}{\pi }\right)+1\right)\right].
\label{eq:1.68frolovfluido}
\end{aligned}
\end{equation}

\begin{equation}
\begin{aligned}
 T^{(\beta\neq2)}=&\left[\sqrt{\pi } l^4 \left((\beta -2) Q \left(\frac{\Lambda  S}{\pi }-3\right)-3 \pi ^{1/\beta } \beta  Q \epsilon  \left(\frac{b}{\sqrt{S}}\right)^{2/\beta }\right)^2\right.
 \\
 &-\frac{9 \sqrt{\pi } (\beta -2)^2 S^2 \left(\frac{S \left(-\pi ^{1/\beta } \epsilon  \left(\frac{b}{\sqrt{S}}\right)^{2/\beta }+\frac{\Lambda  S}{\pi }-1\right)}{\pi }+Q^2\right)}{\pi ^2}
\\
&-\frac{36 \sqrt{\pi } (\beta -2) l^2 Q^2 S \left(\pi ^{1/\beta } (\beta +1) \epsilon  \left(\frac{b}{\sqrt{S}}\right)^{2/\beta }+\beta -2\right)}{\pi }
\\
&\left.-\frac{3 \sqrt{\pi } l^2 S^2 \left((\beta -2) \left(\frac{\Lambda  S}{\pi }-3\right)-3 \pi ^{1/\beta } \beta  \epsilon  \left(\frac{b}{\sqrt{S}}\right)^{2/\beta }\right)^2}{\pi ^2}\right]\div
\\
&36 (\beta -2)^2 S^{3/2} \left(2 l^2 Q^2+\frac{S^2}{\pi ^2}\right)
\label{eq:1.68frolovfluido11}
\end{aligned}
\end{equation}

\begin{figure}[h!]
  \centering
  \begin{subfigure}[b]{0.45\textwidth}
    \includegraphics[width=\textwidth]{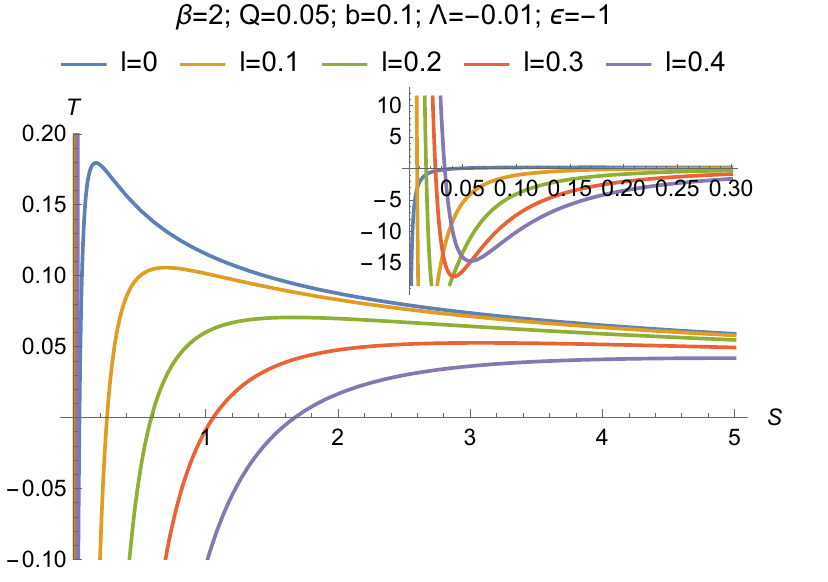}
    \caption{}\label{im3afrolovfluido}
  \end{subfigure}
  \vspace{.5cm}
  \begin{subfigure}[b]{0.47\textwidth}
    \includegraphics[width=\textwidth]{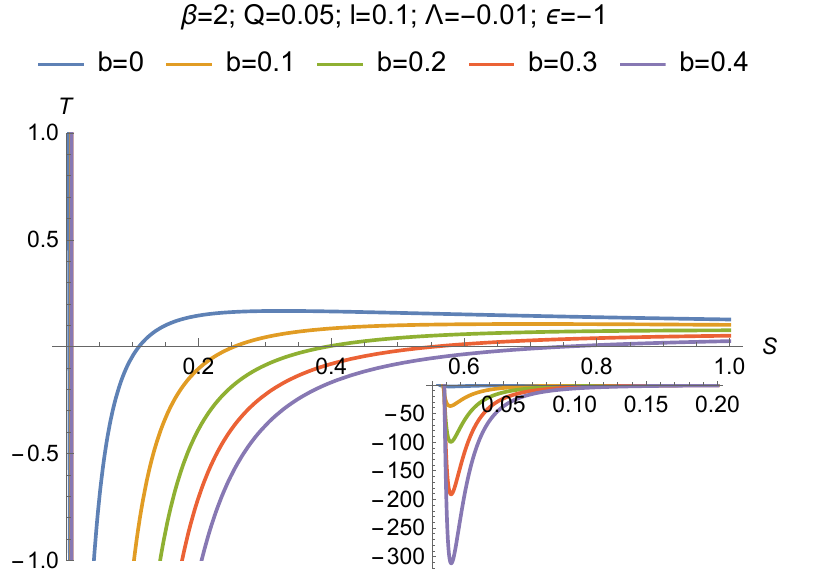}
    \caption{}\label{im3bfrolovfluido}
  \end{subfigure}
   \begin{subfigure}[b]{0.45\textwidth}
    \includegraphics[width=\textwidth]{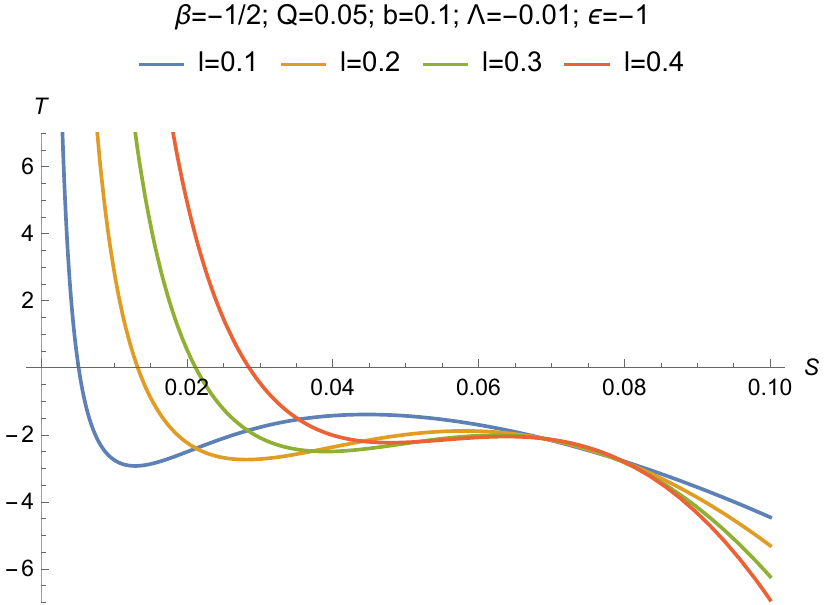}
    \caption{}\label{ffc3}
  \end{subfigure}
   \begin{subfigure}[b]{0.47\textwidth}
    \includegraphics[width=\textwidth]{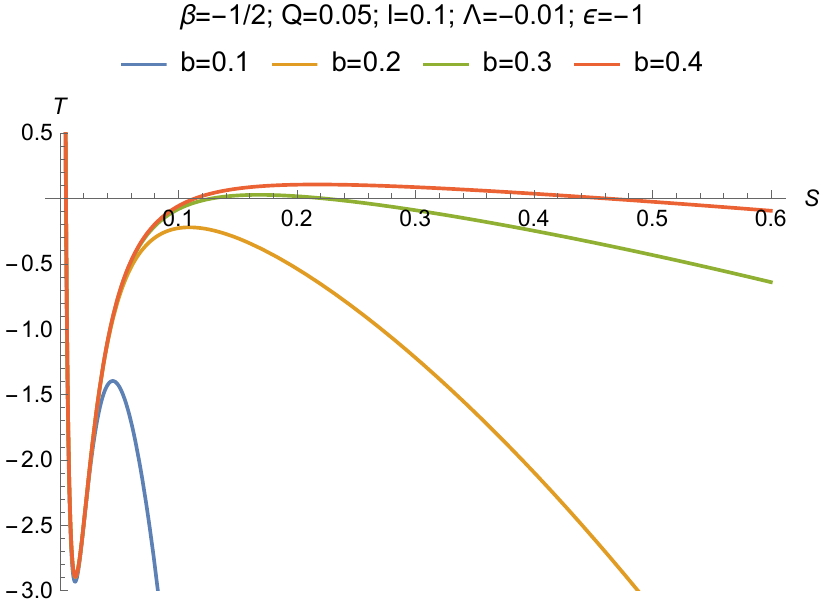}
    \caption{}\label{ffc4}
  \end{subfigure}
 \caption{Black hole temperature as a function of entropy $T(S)$ for different values of $b$ and $l$.}
  \label{imfrolovfluido2}
\end{figure}

In Figs. \ref{im3afrolovfluido}-\ref{ffc4}, we represent the behavior of the temperature parameter, $T$, as a function of the entropy of the black hole, $S$, in different situations. In Fig. \ref{im3afrolovfluido}, you can see that for the Reissner-Nordström-AdS spacetime with fluid of strings ($l = 0$, $Q=0.05$, $\Lambda=-0.01$ and $b=0.1$), the temperature parameter has positive and negative values for $S > 0$. Similarly, for the Frolov-AdS spacetime with fluid of strings ($l\neq 0$, $Q=0.05$, $\Lambda=-0.01$ and $b=0.1$). It can also be seen that for $S=2$ the temperature increases as $l$ decreases.

In Fig. \ref{im3bfrolovfluido}, you can see that for the Frolov-AdS spacetime without fluid of strings ($l = 0.1$, $Q=0.05$, $\Lambda=-0.01$ and $b=0$), the temperature parameter has positive and negative values for
$S > 0$. Similarly, for the Frolov-AdS spacetime with fluid of strings ($l = 0.1$, $Q=0.05$, $\Lambda=-0.01$ and $b\neq0$). It can also be seen that for $S>0.8$, the temperature only shows positive values, increasing as the intensity of the $b$ fluid of strings decreases.

In Figs. \ref{ffc3}-\ref{ffc4}, we illustrate the behavior of the Hawking temperature for regular Frolov-AdS spacetime with fluid of strings ($\beta=-1/2$) for different values of $l$ and $b$. 

It turns out that for the singular Frolov-AdS spacetime with dark matter fluid ($\beta=2$), the limits are opposite to those found in the regular Frolov-AdS spacetime with fluid of strings ($\beta=-1/2$). 

\begin{figure}[h!]
  \centering
  \begin{subfigure}[b]{0.45\textwidth}
    \includegraphics[width=\textwidth]{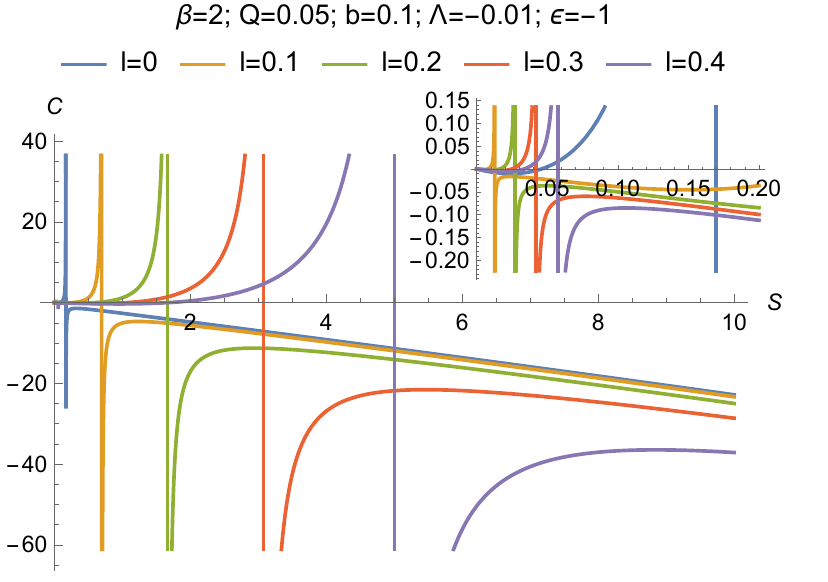}
    \caption{}\label{im4afrolovfluido}
  \end{subfigure}
  \vspace{.5cm}
  \begin{subfigure}[b]{0.47\textwidth}
    \includegraphics[width=\textwidth]{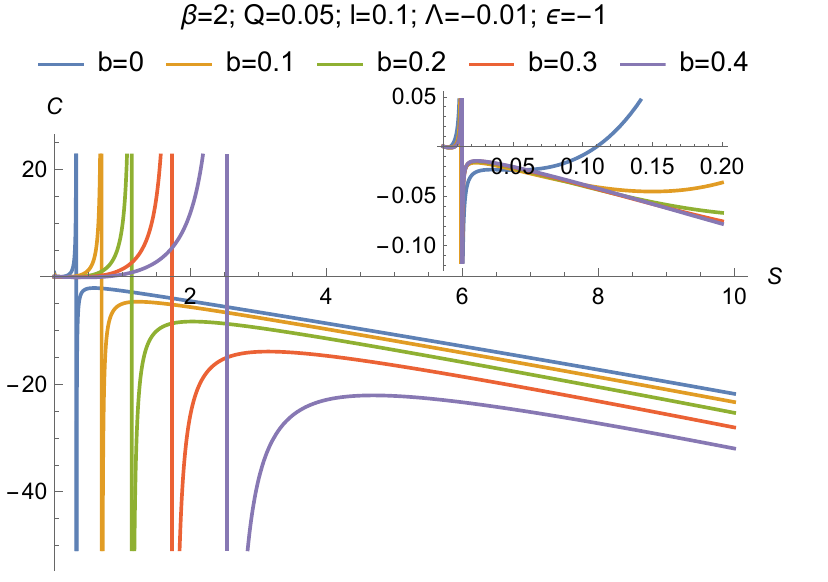}
    \caption{}\label{im4bfrolovfluido}
  \end{subfigure}
   \begin{subfigure}[b]{0.45\textwidth}
    \includegraphics[width=\textwidth]{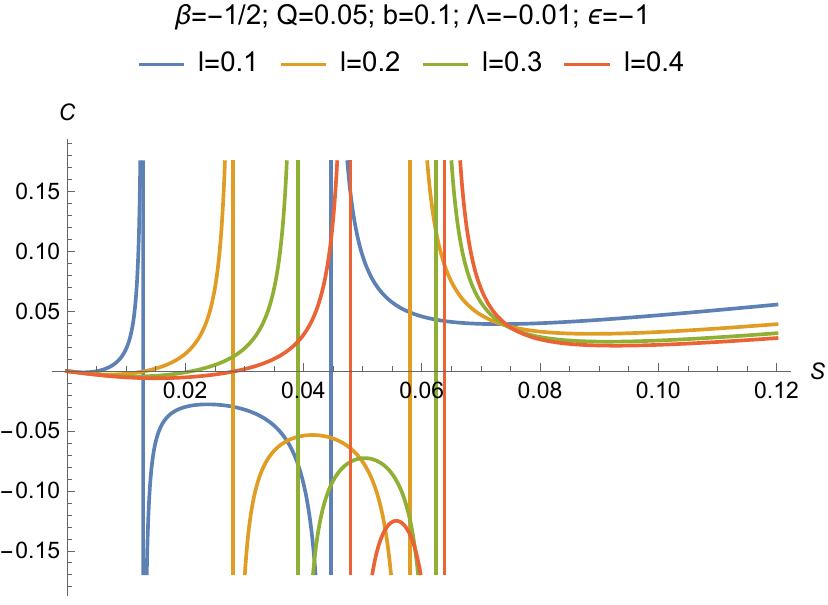}
    \caption{}\label{ffc5}
  \end{subfigure}
   \begin{subfigure}[b]{0.47\textwidth}
    \includegraphics[width=\textwidth]{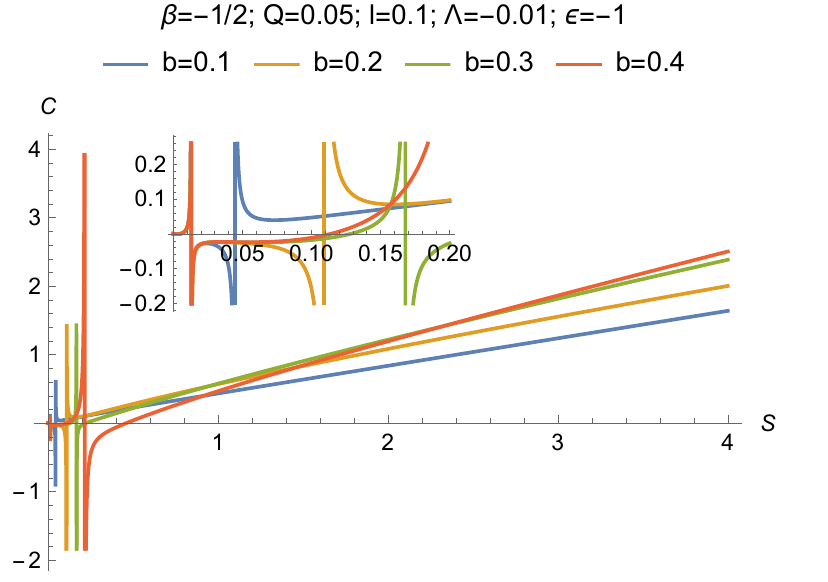}
    \caption{}\label{ffc6}
  \end{subfigure}
 \caption{Heat capacity as a function of entropy $C(S)$ for different values of $b$ and $l$.}
  \label{imfrolovfluido3}
\end{figure}
\subsection{Heat capacity}

In order to calculate the heat capacity of the class of solutions corresponding to  the Frolov-AdS black hole with a fluid of strings, let us use the following relation:

\begin{equation}
C=T\frac{\partial S}{\partial T}=T\left(\frac{\partial T}{\partial S}\right)^{-1}.
\label{eq:1.69frolovfluido}
\end{equation}

Substituting Eq. (\ref{eq:1.68frolovfluido}) into Eq.(\ref{eq:1.69frolovfluido}), we find that the heat capacity as a function of the entropy is given by:
\begin{equation}
\begin{aligned}   
C^{(\beta=2)}=&\left[2 S \left(2 \pi ^2 l^2 Q^2+S^2\right)\right.
\\
&\left.[-36 \pi ^{9/2} b l^4 Q^2 \sqrt{S} \epsilon -36 \pi ^4 l^4 Q^2 S+144 \pi ^3 l^2 Q^2 S^2+108 \pi ^2 l^2 S^3+36 \pi ^2 Q^2 S^3-36 \pi  S^4\right.
\\
&\left.-9 \pi ^5 b^2 l^4 Q^2 \epsilon ^2+36 \pi ^{7/2} b l^2 Q^2 S^{3/2} \epsilon +108 \pi ^{5/2} b l^2 S^{5/2} \epsilon -36 \pi ^{3/2} b S^{7/2} \epsilon\right.
\\
&\left.+27 \pi ^3 b^2 l^2 S^2 \epsilon ^2+12 \pi ^{7/2} b \Lambda  l^4 Q^2 S^{3/2} \epsilon +24 \pi ^3 \Lambda  l^4 Q^2 S^2-72 \pi  \Lambda  l^2 S^4+36 \Lambda  S^5\right.
\\
&\left.-36 \pi ^{3/2} b \Lambda  l^2 S^{7/2} \epsilon -4 \pi ^2 \Lambda ^2 l^4 Q^2 S^3+12 \Lambda ^2 l^2 S^5\right.
\\
&+108 \pi ^3 b^2 l^2 S^2 \epsilon ^2 \log ^2\left(\frac{\sqrt{S}}{\sqrt{\pi }}\right)+24 \pi ^{7/2} b l^2 Q^2 S^{3/2} \epsilon  \left(\Lambda  l^2+9\right) \log \left(\frac{\sqrt{S}}{\sqrt{\pi }}\right)
\\
&\left.-18 \pi ^{3/2} b l^2 \epsilon  \log \left(\frac{S}{\pi }\right) \left(\pi ^3 l^2 Q^2 \left(\sqrt{\pi } b \epsilon +2 \sqrt{S}\right)-3 \pi ^{3/2} b S^2 \epsilon +2 \Lambda  S^{7/2}-6 \pi  S^{5/2}\right)\right.
\\
&\left.-9 \pi ^5 b^2 l^4 Q^2 \epsilon ^2 \log ^2\left(\frac{S}{\pi }\right)]\right]/
\\
&\left[216 \pi ^6 l^6 Q^4 S-288 \pi ^5 l^4 Q^4 S^2+468 \pi ^4 l^4 Q^2 S^3+72 \pi ^4 l^2 Q^4 S^3-936 \pi ^3 l^2 Q^2 S^4\right.
\\
&\left.-324 \pi ^2 l^2 S^5-108 \pi ^2 Q^2 S^5+144 \pi ^{13/2} b l^6 Q^4 \sqrt{S} \epsilon +288 \pi ^{11/2} b l^4 Q^4 S^{3/2} \epsilon +36 \pi  S^6\right.
\\
&\left.+648 \pi ^{9/2} b l^4 Q^2 S^{5/2} \epsilon -144 \pi ^{7/2} b l^2 Q^2 S^{7/2} \epsilon -216 \pi ^{5/2} b l^2 S^{9/2} \epsilon +72 \pi ^{3/2} b S^{11/2} \epsilon\right.
\\
&\left.+18 \pi ^7 b^2 l^6 Q^4 \epsilon ^2+207 \pi ^5 b^2 l^4 Q^2 S^2 \epsilon ^2-27 \pi ^3 b^2 l^2 S^4 \epsilon ^2-48 \pi ^5 \Lambda  l^6 Q^4 S^2\right.
\\
&\left.-336 \pi ^{7/2} b \Lambda  l^4 Q^2 S^{7/2} \epsilon -552 \pi ^3 \Lambda  l^4 Q^2 S^4+360 \pi ^2 \Lambda  l^2 Q^2 S^5+72 \pi  \Lambda  l^2 S^6+36 \Lambda  S^7\right.
\\
&\left.+216 \pi ^5 b^2 l^4 Q^2 S^2 \epsilon ^2 \log \left(\frac{\sqrt{S}}{\sqrt{\pi }}\right)-8 \pi ^4 \Lambda ^2 l^6 Q^4 S^3+132 \pi ^2 \Lambda ^2 l^4 Q^2 S^5+12 \Lambda ^2 l^2 S^7\right.
\\
&\left.-108 \pi ^3 b^2 l^2 S^2 \epsilon ^2 \log ^2\left(\frac{\sqrt{S}}{\sqrt{\pi }}\right) \left(2 \pi ^2 l^2 Q^2+5 S^2\right)\right.
\\
&\left.+6 \pi ^{3/2} b l^2 \epsilon[\right.
\\
&\left.+2 \pi ^4 l^4 Q^4 \left(9 \pi ^{3/2} b \epsilon -4 \Lambda  S^{3/2}+24 \pi  \sqrt{S}\right)\right.
\\
&\left.+3 \pi ^2 l^2 Q^2 S^{3/2} \left(7 \pi ^{3/2} b \sqrt{S} \epsilon -24 \pi ^2 Q^2-12 \Lambda  S^2+16 \pi  S\right)\right.
\\
&\left.-3 S^{7/2} \left(9 \pi ^{3/2} b \sqrt{S} \epsilon +36 \pi ^2 Q^2-4 \Lambda  S^2+24 \pi  S\right)]\log \left(\frac{S}{\pi }\right)\right.
\\
&\left.+9 \pi ^5 b^2 l^4 Q^2 \epsilon ^2 \log ^2\left(\frac{S}{\pi }\right) \left(10 \pi ^2 l^2 Q^2+9 S^2\right)\right]
\label{eq:1.70frolovfluido}
\end{aligned}
\end{equation}

\begin{equation}
\begin{aligned}   
C^{(\beta\neq2)}=\frac{\psi _1-\psi _2-\psi _3+\psi _4}{3 \pi ^2 l^4 Q^2 S \left(\psi _7+\psi _8+\psi _9\right)+3 l^2 S^2 \left(\psi _{10}+\psi _{11}-\psi _{12}\right)+\psi _5+\psi _6}
\label{eq:1.70frolovfluido11}
\end{aligned}
\end{equation}

\begin{equation}
\begin{aligned}   
\psi _1=-\frac{36 \pi  (\beta -2) l^2 Q^2 S \left(\pi ^{1/\beta } (\beta +1) \epsilon  \left(\frac{b}{\sqrt{S}}\right)^{2/\beta }+\beta -2\right)}{\frac{1}{2 \pi ^4 \beta  S \left(2 l^2 Q^2+\frac{S^2}{\pi ^2}\right)}}
\label{psi1}
\end{aligned}
\end{equation}

\begin{equation}
\begin{aligned}   
\psi _2=\frac{9 (\beta -2)^2 S^2 \left(Q^2-\frac{S \left(\pi ^{\frac{1}{\beta }+1} \epsilon  \left(\frac{b}{\sqrt{S}}\right)^{2/\beta }-\Lambda  S+\pi \right)}{\pi ^2}\right)}{\frac{1}{2 \pi ^4 \beta  S \left(2 l^2 Q^2+\frac{S^2}{\pi ^2}\right)}}
\label{psi2}
\end{aligned}
\end{equation}

\begin{equation}
\begin{aligned}   
\psi _3=\frac{3 l^2 S^2 \left((\beta -2) \left(\frac{\Lambda  S}{\pi }-3\right)-3 \pi ^{1/\beta } \beta  \epsilon  \left(\frac{b}{\sqrt{S}}\right)^{2/\beta }\right)^2}{\frac{1}{2 \pi ^4 \beta  S \left(2 l^2 Q^2+\frac{S^2}{\pi ^2}\right)}}
\label{psi3}
\end{aligned}
\end{equation}

\begin{equation}
\begin{aligned}   
\psi _4=\frac{\pi ^2 l^4 \left((\beta -2) Q \left(\frac{\Lambda  S}{\pi }-3\right)-3 \pi ^{1/\beta } \beta  Q \epsilon  \left(\frac{b}{\sqrt{S}}\right)^{2/\beta }\right)^2}{\frac{1}{2 \pi ^4 \beta  S \left(2 l^2 Q^2+\frac{S^2}{\pi ^2}\right)}}
\label{psi4}
\end{aligned}
\end{equation}

\begin{equation}
\begin{aligned}   
\psi _5=9 (\beta -2)^2 S^4 \left(-\pi ^{\frac{1}{\beta }+1} (\beta +2) S \epsilon  \left(\frac{b}{\sqrt{S}}\right)^{2/\beta }+3 \pi ^2 \beta  Q^2-\beta  \Lambda  S^2-\pi  \beta  S\right)
\label{psi5}
\end{aligned}
\end{equation}

\begin{equation}
\begin{aligned}   
\psi _6=&2 \pi ^4 \beta  l^6 Q^4 \left[-9 \pi ^{\frac{2}{\beta }+2} \beta  (3 \beta +4) \epsilon ^2 \left(\frac{b}{\sqrt{S}}\right)^{4/\beta }\right.
\\
&\left.-6 \pi ^{\frac{1}{\beta }+1} (\beta -2) \epsilon  \left(\frac{b}{\sqrt{S}}\right)^{2/\beta } (\pi  (9 \beta +6)\right.
\\
&\left.-(\beta +2) \Lambda  S)-(\beta -2)^2 (3 \pi -\Lambda  S) (\Lambda  S+9 \pi )\right]
\label{psi6}
\end{aligned}
\end{equation}

\begin{equation}
\begin{aligned}   
\psi _7=-3 \pi ^{\frac{2}{\beta }+2} \beta ^2 (13 \beta -20) S \epsilon ^2 \left(\frac{b}{\sqrt{S}}\right)^{4/\beta }
\label{psi7}
\end{aligned}
\end{equation}

\begin{equation}
\begin{aligned}   
\psi _8=&2 \pi ^{\frac{1}{\beta }+1} (\beta -2) \epsilon  \left(\frac{b}{\sqrt{S}}\right)^{2/\beta } [3 \pi  \left(4 \pi  (\beta +1) (\beta +2) Q^2\right.
\\
&\left.+\beta  (10-13 \beta ) S\right)+\beta  (23 \beta -10) \Lambda  S^2]
\label{psi8}
\end{aligned}
\end{equation}

\begin{equation}
\begin{aligned}   
\psi _9=(\beta -2)^2 \beta  \left(24 \pi ^3 Q^2-11 \Lambda ^2 S^3+46 \pi  \Lambda  S^2-39 \pi ^2 S\right)
\label{psi9}
\end{aligned}
\end{equation}

\begin{equation}
\begin{aligned}   
\psi _{10}=9 \pi ^{\frac{2}{\beta }+2} \beta ^2 (3 \beta +4) S^2 \epsilon ^2 \left(\frac{b}{\sqrt{S}}\right)^{4/\beta }
\label{psi10}
\end{aligned}
\end{equation}

\begin{equation}
\begin{aligned}   
\psi _{11}=&6 \pi ^{\frac{1}{\beta }+1} (\beta -2) S \epsilon  \left(\frac{b}{\sqrt{S}}\right)^{2/\beta } \left[\pi ^2 (\beta  (13 \beta +6)+8) Q^2-\beta  (\beta +2) \Lambda  S^2\right.
\\
&\left.+3 \pi  \beta  (3 \beta +2) S\right]
\label{psi11}
\end{aligned}
\end{equation}

\begin{equation}
\begin{aligned}   
\psi _{12}=(\beta -2)^2 \beta  \left(6 \pi ^4 Q^4+3 \pi ^2 S^2 \left(10 \Lambda  Q^2-9\right)-78 \pi ^3 Q^2 S+\Lambda ^2 S^4+6 \pi  \Lambda  S^3\right)
\label{psi12}
\end{aligned}
\end{equation}

From Figs. \ref{im4afrolovfluido}-\ref{ffc6}, we can conclude that there are values of $S$ for which the heat capacity is positive, just as there are values for which the heat capacity is negative. In other words, the black hole can be thermodynamically stable or unstable, and this stability is related to the values of the fluid of strings parameter. We can see that the transition point at which the heat capacity diverges changes when we vary this parameter.

In Figs. \ref{im4afrolovfluido}-\ref{im4bfrolovfluido} the transition point is shifted to the right as we increase the intensities of the parameters $l$ and $b$. For $S>6$, the Frolov-AdS spacetime with dark matter fluid ($\beta=2$) will be thermodynamically unstable, i.e., it will have a negative heat capacity. For this same system, the heat capacity increases as we decrease the parameters $l$ and $b$.

In Figs. \ref{ffc5}-\ref{ffc6} the transition point is shifted to the right as we increase the intensities of the parameters $l$ and $b$. For $S>2$, the regular Frolov-AdS spacetime with fluid of strings ($\beta=-1/2$) will be thermodynamically stable, i.e. it will have a positive heat capacity. For this same system, the heat capacity increases as we decrease the $l$ parameter and increase the $b$ parameter.

\section{Concluding remarks}
\label{sec4} 

 The  class of solutions we have obtained represents a generalization 
 of the original Frolov
regular
black hole solution. This generalization is realized by the addition of the cosmological constant and a fluid of strings, which surrounds the black hole. It is worth calling attention to the fact that the
new 
metrics are significantly different from the standard Frolov
black hole metric
, offering new physical information  about the
geometry and nature of the new class of black holes. Additionally, the obtained metrics
give us 
appropriate  limits, as expected,
confirming the validity of the assumptions we have chosen to obtain the new class of black holes.

For all metrics obtained, an analysis was done in which concerns
the behaviors of the Kretschmann scalar in order to check for the regularity or not of the metrics.
Then, we concluded that the solutions are regular only for values
of $-1\leq\beta<0$. Otherwise, for $\beta =1,2, 3$, the metrics are not regular, as we can see from the fact that the Kretschmann scalar is divergent at the origin.

On the other hand, the analysis of the geodesics with respect to their completeness and incompleteness tells us that 
for $\beta=2$ and $\beta=3$, the geodesics are incomplete, as we can conclude from
Fig. \ref{bfrolovfs}, and for $\beta=-1/10$. $\beta=-1/2$ and $\beta=-1$, the geodesics are complete, as can be read from
Fig. \ref{dfrolovfs}.
Thus, the analysis of the geodesics, for the values of $\beta$ considered, with the aim of knowing the characteristics
of the spacetimes in relation to the singularity, confirms what is predicted by the results provided by the Kretschmann scalar, at least in the cases considered.
This is an interesting and surprising result is that the predictions arising from the analysis of the Kretschmann scalar are confirmed by the analysis of the geodesics.

The investigations related to the black hole thermodynamics, in which concerns to mass in terms of the entropy, Hawking temperature, and heat capacity, were performed  for $\beta = -1/2$ and $\beta=2$, in which cases  
black hole is regular or singular, respectively.  
In both cases, the mass parameter as a function of the entropy has
positive and negative values depending on the parameters of the black hole and on the values of $\beta$.  

The results
related to Hawking temperature and heat capacity are significantly different from the standard ones due to the presence of the cosmological constant and of the fluid of strings, as sources. Note that they strongly depend on the value of $\beta$, as expected, and on values of the parameters characterizing the original Frolov black hole, as well as as on the parameter that codifies the presence of the fluid of strings. These new results offer interesting insights into the subject
concerning the rich geometry of black holes, regular or singular, and their
nature. solutions. 

\backmatter
\bmhead{Acknowledgments}
V. B. Bezerra is partially supported by Conselho Nacional de Desenvolvimento Científico e Tecnológico - CNPq, Brazil, through the Research Project No. 307211/2020-7.

\section*{Declarations}
\textbf{Funding:} Conselho Nacional de Desenvolvimento Científico e Tecnológico - CNPq, Brazil. 

\noindent
\textbf{Conflict of interest/Competing interests:} The authors declare that they have no conflicts of interest.

\noindent
\textbf{Ethics approval and consent to participate:} Not applicable.

\noindent
\textbf{Consent for publication:} The authors consent with publication.

\noindent
\textbf{Data availability:} This study did not utilize underlying data.

\noindent
\textbf{Materials availability:} Not applicable.

\noindent
\textbf{Code availability:} Not applicable.

\noindent
\textbf{Author contribution:}
The authors contributed equally.
\bibliography{ref}
\end{document}